# Generating Concurrency Checks Automatically

Jonathan Hoyland
Royal Holloway, University of London

Matthew Hague
Royal Holloway, University of London

## ABSTRACT

This article introduces ATAB, a tool that automatically generates pairwise reachability checks for action trees. Action trees can be used to study the behaviour of real-world concurrent programs. ATAB encodes pairwise reachability checks into alternating tree automata (ATA) that determine whether an action tree has a schedule where any pair of given points in the program are simultaneously reachable. Because the pairwise reachability problem is undecidable in general ATAB operates under a restricted form of lock-based concurrency. ATAB produces ATA that are more compact and more efficiently checkable than those that have been previously used. The process is entirely automated, which simplifies the process of encoding checks for more complex action trees. The ATA produced are easier to scale to large numbers of locks than previous constructions.

## 1. INTRODUCTION

Analysing programs can give safety guarantees about their behaviour. Programs can be represented by action trees. An action tree represents all the actions taken by a program over the course of its execution as nodes in a tree. Branches in the tree represent different threads in a multi-threaded program. By analysing the action tree of a program it is possible to derive properties of the program.

Action trees have been used to analyse software in a number of contexts. Yasukata et al. [5] analyse a number of Java-like programs using action trees constructed with higher-order recursion schemes (HORS). Nordhoff et al. [3] use action trees to analyse concurrency properties of Java programs, and extend an eclipse plug-in to provide more accurate race-condition detection. The approach is effective because it can analyse very complex programs without having to account for implementation details, examining only the observable behaviour.

If the action tree of a program can be generated by a fixed set of rules then more properties can be determined because of the additional structure in the tree. The more expressive the method used to generate the action tree the more programs can be captured, and thus the more programs can be analysed. This fact, however, is held in tension with the fact that the more expressive the method used to generate the action tree the fewer properties can be decided. Trees with simple constructions are easier to analyse than those with intricate constructions.

In this paper we consider methods for constructing automata that can determine properties of action trees. Specifically we consider a variant of the pairwise reachability problem: given an action tree representing a multi-threaded program, and a list of points in the program, determine whether there is some interleaving of actions of the program such that there are two threads at a listed point at the same time. Although this problem is undecidable under general concurrency, it is decidable (and reasonably expressive) for a restricted subset of concurrency called join lock sensitive (JLS) concurrency. Our construction process is entirely automated, meaning that even for complex trees it is possible to rapidly construct automata that determine the pairwise reachability of the tree.

In Section 6 these automata are used to evaluate action trees constructed using HORS, demonstrating that the automata can be used in practice to determine properties of HORS. We demonstrate that the automata are substantially more compact, efficient, and extensible than those previously used. Furthermore, because the construction is entirely automated, the process is more robust and practical for large examples.

Alternating tree automata (ATA) are an extension of non-deterministic tree automata that can determine many properties of trees including pairwise reachability. Pairwise reachability can be used to determine any number of specific properties through the placement of labels through the tree. For example it is possible to determine if two threads access the same resource simultaneously, or whether two threads become out of step with each other.

ATAB (ATA Builder) is a tool that takes a pairwise reachability problem, and some properties of the action tree and produces an ATA that rejects the action tree if it is pairwise reachable. ATAB translates the pairwise reachability check and the JLS restrictions into a single ATA. Because the ATA varies based on the number of locks and labels in the action tree automating this process makes the onerous and delicate task of encoding the JLS rules and reachability checks for each new action tree much easier.

ATA have previously been constructed to solve the pairwise reachability problem [5]. This paper's contribution is the automation of this construction, along with an improvement to the efficiency of the resulting automata.

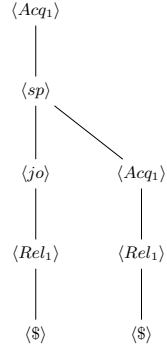

Figure 1: An unschedulable action tree. The child process cannot acquire $Lock_1$ until after the join, but the join cannot complete until after the child process has terminated.

Section 2 defines action trees and action tree forests. Section 3 defines the restrictions needed to make the pairwise reachability problem decidable. Section 4 introduces alternating tree automaton (ATA), and Section 5 introduces ATAB. In Section 6 the automata constructed by ATAB are compared to those used by Yasukata et al who used ATA to determine the pairwise reachability of HORS using HORSAT [1].. The automata produced by ATAB substantially outperform those used by Yasukata, demonstrating their efficacy.

## 2. ACTION TREES

Action trees are a way of expressing the actions taken by a concurrent program. Each branch of the action tree represents a new thread being formed. A non-branching action tree could be considered a trace of a single threaded program.

Action trees are a useful mechanism for analysing programs because they formalise what the program actually does into a form that is easy to reason about. They provide a way to describe any program (written in an arbitrary language) in a form that is easy to analyse. Action trees can be constructed and described in a wide variety of ways, but for the purposes of this document we do not consider their construction until Section 6.

Action trees have four concurrency operators: the $\langle sp \rangle$ operator, which spawns a new thread; the $\langle jo \rangle$ operator, which halts a thread until all of it's children have terminated; and the $\langle acq_x \rangle$ and $\langle rel_x \rangle$ operators, which acquire and release locks respectively.

**Definition 1** (Action Tree). *Action trees are formally defined by the following recursive language:*

$$\gamma ::= \bot \mid \langle \$ \rangle \mid \ell \mid \langle jo \rangle \, \gamma \mid \langle sp \rangle \, \gamma_1 \, \gamma_2 \mid$$
$$\langle Acq_i \rangle \, \gamma \mid \langle Req_i \rangle \, \gamma$$

Here the $\ell$ symbol is the set of all program labels which are used to determine pairwise reachability. $\langle jo \rangle \, \gamma$ is the join operation, with $\gamma$ being the action performed after all child processes have terminated. The spawn operator has two arguments, the continuation of the root process, and the spawned process. The acquire and release symbols are drawn from the set $\{Acq_i, Rel_i \mid i \in [1..k]\}$, where $k$ is the number of locks. The dollar symbol signifies a thread has terminated. The $\bot$ symbol is used as a stand in node for any action that does not affect concurrency.

We define the descendant relation in the usual way: $n_1 \prec n_2$ if the path (starting from the root node) to $n_1$ is a prefix of the path to $n_2$.

**Definition 2** (Action sequence). *An action sequence is a sequence of nodes in the tree, $n_1, n_2, \ldots$, such that*

$$\forall i, j. \ n_i \neq n_j \tag{1}$$
$$\forall i, j. \ n_i \prec n_j \Rightarrow i \leq j \tag{2}$$
$$and \ \forall i \ \forall n \in Tree. \ n \prec n_i \Rightarrow \exists j. \ n_j = n \tag{3}$$

Line 1 ensures nodes are unique. Line 2 ensures actions occur in the correct order. Line 3 ensures that a node can only appear in the action sequence if all its ancestors appear in the sequence also.

An action sequence is said to respect joins if for each thread that has a join, it performs no other actions until all threads it has spawned terminate. An action sequence is said to respect locks if (i) locks are released before they are reacquired, and (ii) locks are acquired before they can be released.

An action sequence is considered well-formed if it respects locks and joins. An action tree is considered well-formed if every branch is a well-formed action sequence.

**Definition 3** (Schedulability). *An action tree is schedulable if there exists some well-formed action sequence that either includes every node in the tree, or is infinitely long i.e. either all threads terminate, or there is always at least one thread that can act.*

It is easy to see that an action tree can be well-formed but unschedulable, for example the tree in Figure 1.

We say nodes $n_i$ and $n_j$ are *simultaneously reachable* if $n_j$ occurs after $n_i$ in an action sequence, but before any of $n_i$'s children. That is, given a well-formed action sequence, $\mathcal{S} := [n_1, n_2, \ldots]$, a node, $n_i \in \mathcal{S}$, with children, $Children_i$, and a second node, $n_j \in \mathcal{S}$ such that $i < j$, we say $n_i$ and $n_j$ are *simultaneously reachable* if $\forall n_k \in Children_i \cap \mathcal{S}. \ j < k$. Solving the pairwise reachability problem involves checking there is some schedule such that a pair of labels is reached simultaneously. However the various checks that together determine schedulability do so by checking that the two relevant threads in the tree reach their final node simultaneously.

To examine properties of threads whilst they are still running we mark the points of interest with *labels*. By terminating threads at labels the checks detect whether a run has a schedule that reaches those labelled, now final, states simultaneously. However because labels could occur at multiple points in a single thread and furthermore the behaviour of a thread after a label may affect the reachability of other labels we must analyse each pair of labels individually. To do this, labels are treated as non-deterministic termination.

A *forest of action trees* is created such that for each pair of labels, there is a tree that terminates when they are reached, and ignores all other labels. More specifically, for each pair of labels, $l_1, l_2$, such that the path to $l_1$ is not a prefix of the path to $l_2$ and vice versa, a copy of the action tree is created where the threads containing the labels are truncated at the label, but all other threads remain unchanged. Such a tree is detected as pairwise reachable if there exists some schedule such that the two labels are simultaneously reachable. This forest of trees is then joined into one larger tree by means of a special $\langle br \rangle$ operator. A second forest is then grown by

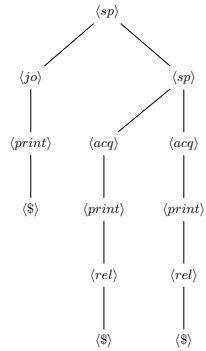

2: An action tree that accesses a printer.

creating a copy of the first forest for each node labelled by a single-child concurrency operator, i.e. *Acq*, *Rel*, or *Jo*. For each copy the node in question is replaced with a (terminating) $\bot$. This larger forest is then joined into one larger tree, again with the $\langle br \rangle$ operator [1]. This second forest is used to aid in the evaluation of infinite branches, as discussed in Section 4. The larger tree is then evaluated by taking the conjunction of all the pairwise checks for each action tree, thus rejecting the entire tree if any pair of labels is reachable.

As expanded later, ATAB constructs automata that take a (restricted) forest of well-formed action trees, and check whether any of the action trees has a schedule that that is pairwise reachable.

To illustrate this process, consider an example from Gawlitza et al.'s paper.

**Example 1** (From [2, Example 1]). Consider a program which spawns two threads and then performs a join. Each of the spawned threads acquires a lock on a printer, prints something and terminates. The root thread then prints something and terminates, without acquiring the printer lock.

The action tree for this is example is shown in Figure 2. To determine whether the printer could ever be accessed by two threads simultaneously, place a label at each instance of the $\langle print \rangle$ action. Now construct a forest that has an action tree for each pair of labels, joined into a tree by $\langle br \rangle$ as in Figure 3[2] If any of these subtrees is pairwise reachable then the printer can be accessed simultaneously by two different threads, causing a clash.

## 3. CONCURRENCY

JLS concurrency is a restriction on full concurrency that allows for dynamic thread creation and termination, and nested use of locks. Informally, nested locking is where a thread must always release its most recently acquired lock before any others it may hold. This pattern of lock acquisitions and releases is called a well-bracketing, so called because it describes the pattern formed by brackets in their usual ordering, e.g. ([ ]) is well-bracketed, but ([ )] is not. Dynamic thread creation and termination consists of *spawn* operations, which create a new child thread[3], and *join* operations, which stall a thread until all of its children have terminated.

This particular pattern of concurrency is notable because Gawlitza et al. [2] proved that the schedulability of JLS action trees is decidable with a regular tree automaton, which is not true of general action trees. Gawlitza et al. used this to decide the pairwise reachability of JLS dynamic pushdown networks (DPNs). This work was latter extended by Yasukata et al. [5] to decide the pairwise reachability of JLS HORS. For any type of JLS tree constructor, if it is possible to decide whether the trees generated are an element of a regular language then it is possible to decide the pairwise reachability property of such tree constructors.

**Definition 4** (Pairwise reachability). *Given an action sequence, $\mathcal{S}$, and a set of labels, Lab, $\mathcal{S}$ is pairwise reachable if it has two threads, $u_1$, $u_2$, that terminate in labels, $l_1, l_2 \in Lab$, such that $u_1 \neq u_2$. The pairwise reachability problem is the problem of deciding, given an action tree and a set of labels, whether there is a schedulable action sequence with two such paths.*

**Lemma 1** (Corollary of [5, Theorem 2]). *The pairwise reachability of a JLS action tree can be determined with a regular language.*

*Proof.* JLS schedulability can be expressed as a regular tree language [2]. Given an action tree, and two labels, $l_1$ and $l_2$, it is possible to determine using a regular tree automaton if the two labels are simultaneously reachable. As regular tree languages are closed under intersection, the intersection of an automaton determining JLS schedulability with automata determining reachability for each pair of labels yields another regular tree automaton. Thus the pairwise reachability of a JLS action tree can be determined with a regular language. □

An action tree is considered *join-lock schedulable* if there is at least one ordering of actions that can be run to completion that respects (nested) locks and joins. The set of programs that respect locks and joins is not simply the intersection of programs that respect locks and those that respect joins. A program may be lock-schedulable (i.e. have a run that satisfies the nested locking properties) and be join-schedulable (i.e. have a run that satisfies the spawn and join rules) but have no schedule that satisfies both properties simultaneously. Furthermore an analysis not sensitive to joins may be unable to find a lock-schedule, even if one exists, because the schedule relies on the communication that occurs via the join. This is equally true of an analysis not sensitive to locks.

Let us reconsider the printer example, Example 1. A join insensitive analysis would report a possible violation between the root thread and a child thread, being unaware that the root thread must hang until all its children have terminated. A lock insensitive analysis would also spuriously report violations, finding a possible violation between the two child threads.

We now formally define some of the concepts from the preceeding paragraph. A *locking sequence* is a thread eliding all non-locking actions. *Nested locking* is when a locking sequence, given a finite number of locks $\{Lock_1, \ldots, Lock_k\}$ with the corresponding release and acquires $\{Acq_i, Rel_i \mid i \in [1..k]\}$, is a prefix of the grammar[4]:

$$L \rightarrow \epsilon \mid L\ L \mid Acq_1\ L\ Rel_1 \mid \cdots \mid Acq_k\ L\ Rel_k$$

---

[1] For a more detailed explanation of this process see Appendix D.

[2] The full construction would also have extra branches for each single-child concurrency operator, but as this tree has no infinite threads we can leave them out for reasons of space safely.

[3] Locks held by a parent thread do not pass to the child, they remain with the parent thread.

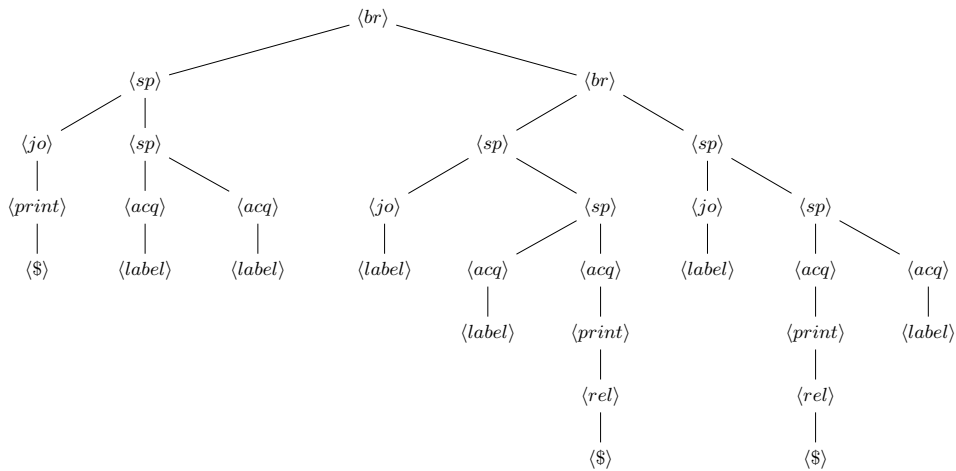

Figure 3: An action forest

The locking sequence is also required to respect locks. An action sequence whose locking sequence has both these properties is referred to as a *lock-well-formed action sequence*. A process is considered *join-lock-well-formed* if (i) it is lock-well-formed, (ii) there are no actions performed after the termination action, $, and (iii) if a branch terminates in the $ action then the corresponding locking sequence is $\in L$, i.e. all locks acquired during the sequence are released before termination. A join-lock-tree is join-lock well-formed if all branches are join-lock well-formed. This is a restriction on well-formed action trees, as defined earlier. ATAB constructs automata that operate over forests of join-lock-well-formed action trees, and determine if there is an action tree in the forest that has a pairwise reachable schedule.

Most model checking requirements are expressed in terms of safety properties, rejecting models for which there exists a path to a state that violates some property. Because regular tree automata are closed under intersection and complementation, building an automaton that checks the pairwise reachability of JLS action trees can be achieved by taking a JLS schedulability automaton, $L_{sched}$, and a pairwise reachability automaton, $L_1$, and constructing the automaton $\neg(L_{sched} \cap \neg L_1)$. This automaton rejects action trees with a schedulable run that violates $L_1$. In practice we use automata that take action forests, and simply take the conjunction of the results of the respective action trees.

## 4. ALTERNATING TREE AUTOMATA

To determine whether the tree holds a property $\varphi$ an automaton can be used. There are any number of different automaton constructions, each with advantages and disadvantages. Alternating tree automata (ATA) are an extension of non-deterministic tree automata. When a non-deterministic tree automaton reaches a choice, if any of the options are accepting, it accepts. This is akin to a disjunction. ATA on reaching a choice can evaluate any boolen formula over the options, or alternatively they can alternate between disjunction and conjunction. For example a non-deterministic tree automaton might have the rule $q_1\ a \rightarrow q_1 \vee q_2 \vee q_3$, whereas an ATA could have the rule $q_1\ a \rightarrow q_1 \wedge (q_2 \vee q_3)$. Although regular tree automata are closed under complementation, and thus non-deterministic tree automata and ATA are equi-expressive, this is a useful extension because the tranformation from ATA to non-deterministic tree automata is potentially exponential. The automata needed to decide pairwise reachability can have dozens of states when expressed as ATA, thus using ATA is more efficient.

Formally an ATA is a four-tuple $\mathcal{A} \coloneqq \langle \Sigma, Q, \delta, q_1 \rangle$ where $\Sigma$ is a ranked alphabet of symbols and their arities, $Q$ is a finite set of states, $q_1 \in Q$ is designated the start state, and $\delta : Q \times \Sigma \rightarrow \mathcal{B}^+(\{1..m\} \times Q)$ is a transition function where $m$ is the arity of $x \in \Sigma$ and $\mathcal{B}^+(X)$ is the set of boolean formulas over $X$.

A run of an ATA, $\mathcal{A}$, on a tree, $t$, is informally defined as a traversal of $t$, where for each terminal in $t$ there is some rule in $\delta$ that moves the current state of $\mathcal{A}$ to some boolean formula, that is evaluated on the terminals children. A run is accepting if it meets the acceptance criteria of the ATA.

An ATA can have a number of different acceptance conditions, but in this document the trivial Büchi acceptance condition is used. Because we are using ATA, and thus are evaluating boolean formulae, a run is accepting if the initial state evaluates to true. States can evaluate to true in two ways, if the boolean formula of the rule triggered by the next element of the input tree is true, or if the state is visited infinitely often whilst traversing the input tree. This is equivalent to a Büchi condition where all states are accepting. In our case, trees that would inappropriately return true because they are infinite, rather than because they match the property, are dealt with by checking a forest of finite prefixes of the tree, as constructed in Section 2. This forest is guaranteed to contain a finite prefix of the tree such that the property is correctly determined, as all of these properties are safety properties, and thus if they exist, they must exist after a finite prefix.

ATA can easily be composed together to check multi-ple properties; solving the pairwise reachability efficiently requires this. To solve the pairwise reachability problem, one must check that the labels are simultaneously reachable, and that the program is schedulable. ATAB is the only fully automated tool for constructing these automata, and produces more compact automata for large numbers of locks than those used by Yasukata et al. [5]

---

[4]Due to the lock-well-formed-ness requirement this grammar is actually a regular language, as it has a finite nesting depth.

## 5. ATAB

ATAB is a tool that takes as *input* (i) the number of locks, (ii) the number of labels, and (iii) and a list of pairs of labels that are to be checked.

ATAB *outputs* an ATA that

- **assumes** that the tree it is to consume is a forest of join-lock-well-formed action trees.

- **checks** that no pair of labels given in the input is pairwise reachable.

The tool can generate the necessary checks depending on the number of labels and locks. Schedulability is determined using a variant of the algorithm presented by Gawlitza et al. [2]. There are three properties that together determine whether a tree has a schedule: (i) double final acquisition, whether a lock is acquired and never released by more than one thread; (ii) child termination, whether all children of thread terminate in the case of a join; and (iii) deadlock detection, whether there is some condition such that no thread can advance. These properties, and the automata that decide them are discussed at length in Appendices A and C, and a sample automaton is included in Appendix B. These automata are included for completeness but do not differ substantially from those used by Yasukata et al [5].

If all three properties are unsatisfied then a schedule exists for the actions in the tree. This simple disjunction works because double final acquisition and child termination are properties of the action tree rather than an action sequence. That is to say that if any (join-lock-well-formed) action sequence of an action tree has one of these properties then all action sequences do. Deadlock detection, on the other hand, is a property that can be true on some schedules of an action tree, but false on others. Consider an action tree for which half its schedules have a deadlock and the other half have a double final acquistion. Such a tree would be found safe, even though it is not detected as unschedulable by the deadlock detection automaton, see Subsection 5.1, because if half the schedules have a double final acquisition then all the schedules do, and thus all the action sequences are found to be unschedulable by the double final acquisition automaton. Pairwise reachability is also a property of the action tree. Thus the full ATA is formed of the disjunction of the three properties disjuncted with the pairwise reachability property, i.e. a forest of action trees is safe if for each action tree (i) all sequences deadlock, (ii) any sequence has (and thus all sequences have) a double final acquisition, (iii) any sequence has (and thus all sequences have) non-terminating children, or (iv) any sequence is not (and thus all sequences are not) pairwise reachable.

## 5.1 Deadlock Detection

Deadlock detection is the most complex of the three schedulability properties to verify, because deadlock can happen in two different ways. The first is where a child thread is waiting for a lock to be released before terminating, but the parent thread won't release the lock until the child thread terminates. The second case is if there is a cycle of lock acquisitions, where a group of threads are all waiting on another to continue before continuing themselves. Because we are concerned with schedulablility, we only consider deadlocking trees unschedulable if all schedules deadlock. However, it is worth noting that there may be action trees that have a deadlocking schedule that are not unschedulable[5].

### 5.1.1 Cycle Detection

Detecting cycles is the most complex property determined by the schedulability automaton. The check is built from a number of instances of the widget shown in Fig 4. Fig 4 determines if $Lock_x$ depends on $Lock_y$. $Lock_x$ depends on $Lock_y$ if $Lock_x$ directly depends on $Lock_y$ or $Lock_x$ indirectly depends on $Lock_y$. $Lock_x$ directly depends on $Lock_y$ if $Lock_y$ is acquired after $Lock_x$ is finally acquired. $Lock_x$ indirectly depends on $Lock_y$ if $Lock_x$ depends directly on $Lock_z$ and $Lock_z$ depends on $Lock_y$. To prevent infinite search paths it is important that direct dependence checks are performed before indirectly dependence checks. From this dependency check a cycle check can be built. There exists a cyclic dependency if there is a lock that depends on itself.

The automaton in Fig 4 detects whether $Lock_x$ depends on $Lock_y$. In the diagram, hollow boxes are used to indicate disjunction, and filled boxes to indicate conjunction. Starting at state $q_{lock_x \to lock_y}$ the automaton loops until $Lock_x$ is acquired. When $Lock_x$ is acquired the automaton guesses whether the lock will ever be released. If it guesses that it will not it continues down the action tree and confirms that the lock is never released, aborting if it is. If the lock is never released, i.e. it is finally acquired, the automaton then guesses whether $Lock_x$ depends on $Lock_y$ directly or indirectly. If it guesses the dependency is direct, it checks whether $Lock_y$ is acquired at some point further down the tree. If it isn't acquired then the automaton aborts. If the automaton instead guesses that the dependency is direct it guesses which lock is next in the chain. In Fig 4 only $Lock_z$ is considered, for reasons of space. Assuming that $Lock_z$ has been guessed, the automaton runs two checks on the remainder of the tree. First it checks that $Lock_z$ is indeed acquired at some future point, aborting if not. Second it checks that $Lock_z$ is dependent on $Lock_y$. If both these checks are positive, then $Lock_x$ depends on $Lock_y$.

## 5.2 Example

Consider again the printer example, Example 1. Consider each of the three action trees in the action forest, shown in Fig 3, individually. The left-most is unschedulable, as there is a double final acquisition, both of the spawned threads acquire the lock and never release it. The central action tree has a deadlock, as the central thread never terminates, and thus the main thread cannot proceed beyond the join, resulting in deadlock. The right-most action tree is analogous to the central action tree. Thus none of the action trees is pairwise reachable, and the action tree forest is safe.

---

[5]For example the tree
$sp\ (acq_1(acq_2(rel_2(rel_1\ \$))))(acq_2(acq_1(rel_1(rel_2\ \$))))$ can deadlock, but would not be detected as unschedulable, as there is a schedule that does not deadlock.

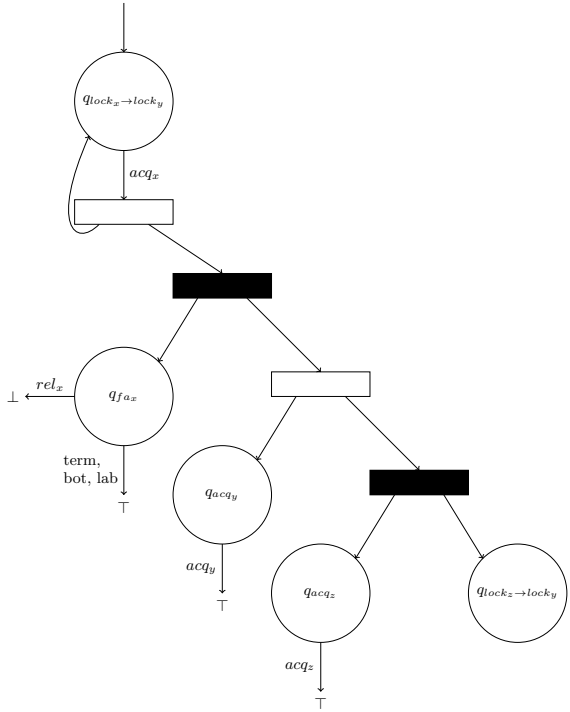

Figure 4: Dependency widget

## 6. RESULTS AND CONCLUSIONS

ATAB can produce ATA for an arbitrary number of locks. Yasukata's automata suffer from a lack of pruning impossible runs early on, and due to the construction of the cyclic-dependency check are difficult to scale to a large number of locks. Yasukata's automata have direct dependency checks for each pair of locks, which are used to check for cycles by manually enumerating every possible cycle of dependencies at the top level. ATAB constructs indirect dependency checks for each pair of locks, and at the top level simply checks each lock to see if it depends on itself. Because the indirect dependency check detects all the possible cyclic paths there are many fewer checks at the top level. Furthermore, because the dependency check checks for direct dependency before checking for indirect dependency there is much less wasted computation checking for longer cycles, when shorter cycles exist.

Fig 5 compares the speed of checking Yasukata's ATA against checking those produced by ATAB. The ATA produced by ATAB can be checked substantially more quickly than those produced by Yasukata. The only check for which Yasukata's ATA performs better than the one produced by ATAB is 'exception_wrong.hors'. On further investigation it transpires that this is because Yasukata's automaton accidentally elides the cyclic-dependency check. However, because in this example there is no cycle, the check appears to pass correctly.

ATAB makes it substantially easier to check HORS for the pairwise reachability property, and produces better automata than those used by Yasukata. It also makes it easier to construct automata for higher numbers of locks with a more compact and efficient way of checking for inter-lock dependencies. The automation of automaton construction also helps prevent mistakes from slipping in, as they did in Yasukata's 'exception_wrong.hors' check.

|  | Yasukata (secs) | ATAB (secs) |
|---|---|---|
| example.hors | 0.43 | 0.13 |
| example_wrong.hors | 0.38 | 0.16 |
| exception.hors | 1.14 | 0.51 |
| exception_wrong.hors | 0.09 | 0.17 |
| list11.hors | 2.36 | 0.15 |
| list12.hors | 3.66 | 0.17 |
| sync11.hors | 3.32 | 1.02 |
| sync12.hors | 4.31 | 1.27 |

Figure 5: Benchmarks

Although Yasukata shows the problem of model checking JLS HORS is decidable, it is not an efficient process. The problem is exponential, for an order-$k$ HORS the problem is $k$-EXPTIME [4]. This may seem like an impossibly expensive algorithm to run on real-world examples, however, as Yasukata notes, if the number of locks and the order of the HORS are fixed then the algorithm is linear in the size of the program. This is a very useful result because it means that even for large programs, if they have relatively few locks, and low order functions then model checking is plausible. Many programs have relatively small requirements in this regard, thus if it is possible to check toy examples with a given tool, with some effort it should be possible to analyse any program of that order with the tool.

Further work on automating the construction of ATA that check different properties including computation tree logic (CTL) is ongoing. A copy of ATAB is available at https://bitbucket.org/jh

## 7. ACKNOWLEDGMENTS

This work was supported by the Engineering and Physical Sciences Research Council [EP/K035584/1 and EP/K009907/1]

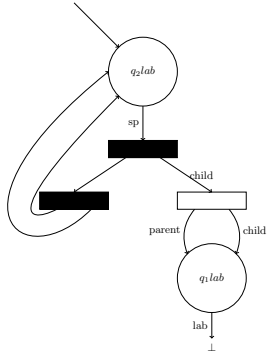

Figure 6: Pairwise reachability widget

# APPENDIX
## A. WIDGETS

The widgets used to decide the schedulability of an action tree are included here for completeness. The widgets are based on those used by Yasukata et al. in [5].

### A.1 Pairwise Reachability

To check pairwise reachability the tool generates a widget for each label that checks that no label terminal appears on more than one branch. Specifically it checks at each spawn whether both the parent and child threads contain a label. Because labels truncate the remainder of the thread, if a label can be reached on both branches it can be reached on both branches simultaneously.

Fig 6 shows a simplified version of the pairwise reachability widget. The widget rejects trees where the parent and child of a spawn both have a label. When put together with the schedulability widgets the automata will reject only trees that are both schedulable and pairwise reachable. In the figure state $q_1 lab$ is only false when both the parent and child thread are labeled, and thus only trees with a path with this property are rejected. At a spawn there are three possible ways for a path to fulfill this property. Either the parent thread will have two instances of the label, the child thread will have two instances of the label, or both the parent and child have one instance of the label.

### A.2 Double Final Acquisition

A lock is said to be finally acquired if it a thread holding the lock terminates without releasing it. If a lock is finally acquired on two different threads then the tree is unschedualable because whichever thread finally acquires the lock first prevents the second thread from ever acquiring it.

Fig 7 shows a simplified version of the double final acquisition widget. At each spawn the automata guesses if the lock is finally acquired on both the parent and child thread. If this is the case the automata accepts, because the tree is unschedualable. The $qr*$ states indicate the lock is released, the $qa*$ states that it has been acquired. The number, e.g. the 2 in $qa_2 lock$ refers to the number of times the lock is guessed to have been acquired. Because the ATA is a top-down automaton it cannot know in advance how many times a lock has been acquired, and thus guesses at each spawn that it has been finally acquired twice. When a spawn happens whilst in the state $qa_2 lock$ the child thread enters the $qr_2 lock$ state. This is beacause only the parent thread keeps

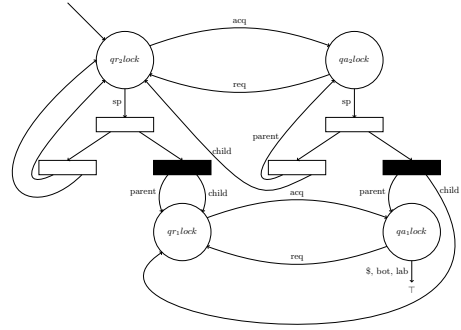

Figure 7: Double final acquisition widget

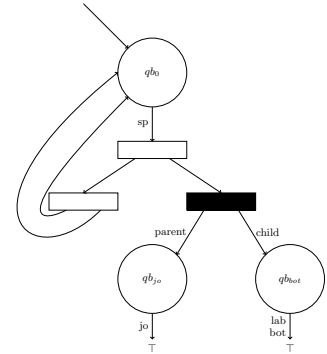

Figure 8: Child termination widget

the lock.

### A.3 Child Termination

Child termination simply enforces the join primitive. At a spawn, the automata guesses if the parent thread will contain a join, and if so ensures termination of the child thread. If there is a join on the parent thread and a label or a $\bot$ on the child thread then the automata accepts because the program is unschedualable.

### A.4 Join-Lock Dependence

Detecting a join-lock dependence is relatively simple. If a thread holding a lock, represented by state $qb_a lock$, is followed by a spawn then the automaton guesses if there is a join on the parent thread. If there is a join on the parent thread before a release, and the child thread or a child of the child thread uses the lock in question then the automaton accepts the tree as unschedualable. When determining if the childs children use the lock, the child must have a join, ensuring that its children finish to ensure the behaviour of the join is respected. Otherwise a child could terminate leaving a grandchild thread running, leading the parents join to appear satisfied erroneously.

## B. SAMPLE ALTERNATING TREE AUTOMATON

This is an automaton generated by ATAB that checks whether $label_1$ and $label_2$ are reachable for a two lock action tree.

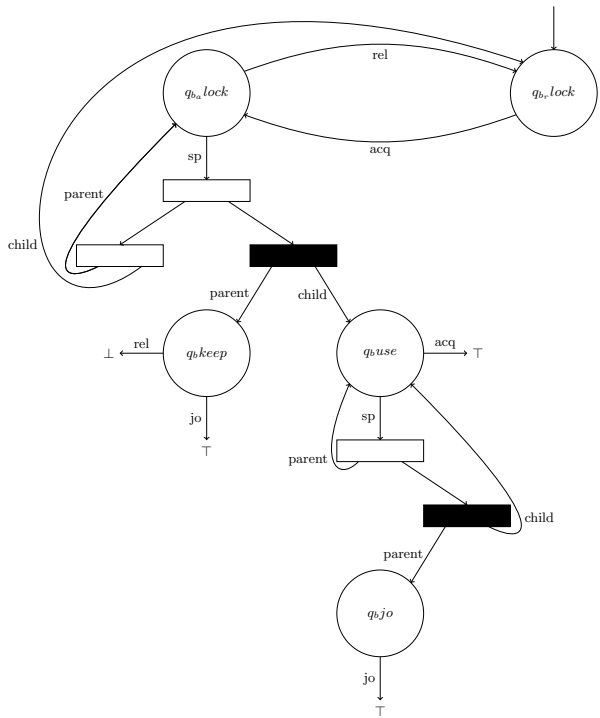

Figure 9: Join-lock dependence widget

```
q0 acq_1 -> ((1, q1_label1) \/ (1, q1_label2) \/ ((1,
    qa2_lock1) \/ (1, qb_a_lock1)) \/ ((1, qr2_lock2)
    \/ (1, qb_r_lock2)) \/ (1, qb_0) \/ ((1,
    q_tca_lock1_lock1) \/ (1, q_tca_lock2_lock2)) \/
    ((1, q_tca_lock1_lock1) \/ ((1, q_fa_lock1) /\ ((1,
     q_acq_lock1) \/ ((1, q_acq_lock2) /\ (1,
    q_tca_lock2_lock1)))))) \/ ((1, q_tca_lock1_lock2)
    \/ ((1, q_fa_lock1) /\ (1, q_acq_lock2)))).
q0 rel_1 -> ((1, q1_label1) \/ (1, q1_label2) \/ ((1,
    qr2_lock2) \/ (1, qb_r_lock2)) \/ (1, qb_r_lock1)
    \/ (1, qb_0) \/ ((1, q_tca_lock1_lock1) \/ (1,
    q_tca_lock2_lock2))).
q0 acq_2 -> ((1, q1_label1) \/ (1, q1_label2) \/ ((1,
    qr2_lock1) \/ (1, qb_r_lock1)) \/ ((1, qa2_lock2)
    \/ (1, qb_a_lock2)) \/ (1, qb_0) \/ ((1,
    q_tca_lock1_lock1) \/ (1, q_tca_lock2_lock2)) \/
    ((1, q_tca_lock2_lock1) \/ ((1, q_fa_lock2) /\ (1,
    q_acq_lock1))) \/ ((1, q_tca_lock2_lock2) \/ ((1,
    q_fa_lock2) /\ ((1, q_acq_lock2) \/ ((1,
    q_acq_lock1) /\ (1, q_tca_lock1_lock2))))))).
q0 rel_2 -> ((1, q1_label1) \/ (1, q1_label2) \/ ((1,
    qr2_lock1) \/ (1, qb_r_lock1)) \/ (1, qb_r_lock2)
    \/ (1, qb_0) \/ ((1, q_tca_lock1_lock1) \/ (1,
    q_tca_lock2_lock2))).
q0 label_1 -> true.
q0 label_2 -> true.
q0 sp -> ((1, q1_label1) /\ (2, q1_label1) \/ ((1,
    q1_label2) /\ (2, q1_label2)) \/ ((1, qr2_lock1) \/
    (2, qr2_lock1) \/ ((1, qr1_lock1) /\ (2, qr1_lock1
    ))) \/ ((1, qb_r_lock1) \/ (2, qb_r_lock1)) \/ ((1,
    qr2_lock2) \/ (2, qr2_lock2) \/ ((1, qr1_lock2) /\
    (2, qr1_lock2))) \/ ((1, qb_r_lock2) \/ (2,
    qb_r_lock2)) \/ ((1, qb_0) \/ (2, qb_0) \/ ((1,
    qb_jo) /\ (2, qb_bot)))) \/ ((1, q_tca_lock1_lock1)
     \/ (1, q_tca_lock2_lock2)) \/ ((2,
    q_tca_lock1_lock1) \/ (2, q_tca_lock2_lock2))).
q0 jo -> ((1, q1_label1) \/ (1, q1_label2) \/ ((1,
    qr2_lock1) \/ (1, qb_r_lock1)) \/ ((1, qr2_lock2)
    \/ (1, qb_r_lock2)) \/ (1, qb_0) \/ ((1,
    q_tca_lock1_lock1) \/ (1, q_tca_lock2_lock2))).
q0 br -> ((1, q0) /\ (2, q0)).
q0 term -> true.
q0 bot -> true.

q1_label1 acq_1 -> (1, q1_label1).
q1_label1 rel_1 -> (1, q1_label1).
q1_label1 acq_2 -> (1, q1_label1).
q1_label1 rel_2 -> (1, q1_label1).
q1_label1 label_1 -> false.
q1_label1 label_2 -> true.
q1_label1 sp -> ((1, q1_label1) /\ (2, q1_label1)).
q1_label1 jo -> (1, q1_label1).
q1_label1 br -> true.
q1_label1 term -> true.
q1_label1 bot -> true.

q1_label2 acq_1 -> (1, q1_label2).
q1_label2 rel_1 -> (1, q1_label2).
q1_label2 acq_2 -> (1, q1_label2).
q1_label2 rel_2 -> (1, q1_label2).
q1_label2 label_1 -> true.
q1_label2 label_2 -> false.
q1_label2 sp -> ((1, q1_label2) /\ (2, q1_label2)).
q1_label2 jo -> (1, q1_label2).
q1_label2 br -> true.
q1_label2 term -> true.
q1_label2 bot -> true.

qr2_lock1 acq_1 -> (1, qa2_lock1).
qr2_lock1 rel_1 -> false.
qr2_lock1 acq_2 -> (1, qr2_lock1).
qr2_lock1 rel_2 -> (1, qr2_lock1).
qr2_lock1 label_1 -> false.
```

```
qr2_lock1 label_2 -> false.
qr2_lock1 sp -> ((1, qr2_lock1) \/ (2, qr2_lock1) \/
    ((1, qr1_lock1) /\ (2, qr1_lock1))).
qr2_lock1 jo -> (1, qr2_lock1).
qr2_lock1 br -> true.
qr2_lock1 term -> false.
qr2_lock1 bot -> false.

qr1_lock1 acq_1 -> (1, qa1_lock1).
qr1_lock1 rel_1 -> false.
qr1_lock1 acq_2 -> (1, qr1_lock1).
qr1_lock1 rel_2 -> (1, qr1_lock1).
qr1_lock1 label_1 -> false.
qr1_lock1 label_2 -> false.
qr1_lock1 sp -> ((1, qr1_lock1) \/ (2, qr1_lock1)).
qr1_lock1 jo -> (1, qr1_lock1).
qr1_lock1 br -> true.
qr1_lock1 term -> false.
qr1_lock1 bot -> false.

qa2_lock1 acq_1 -> false.
qa2_lock1 rel_1 -> (1, qr2_lock1).
qa2_lock1 acq_2 -> (1, qa2_lock1).
qa2_lock1 rel_2 -> (1, qa2_lock1).
qa2_lock1 label_1 -> false.
qa2_lock1 label_2 -> false.
qa2_lock1 sp -> ((1, qa2_lock1) \/ (2, qr2_lock1) \/
    ((1, qa1_lock1) /\ (2, qr1_lock1))).
qa2_lock1 jo -> (1, qa2_lock1).
qa2_lock1 br -> true.
qa2_lock1 term -> false.
qa2_lock1 bot -> false.

qa1_lock1 acq_1 -> false.
qa1_lock1 rel_1 -> (1, qr1_lock1).
qa1_lock1 acq_2 -> (1, qa1_lock1).
qa1_lock1 rel_2 -> (1, qa1_lock1).
qa1_lock1 label_1 -> true.
qa1_lock1 label_2 -> true.
qa1_lock1 sp -> ((1, qa1_lock1) \/ (2, qr1_lock1)).
qa1_lock1 jo -> (1, qa1_lock1).
qa1_lock1 br -> false.
qa1_lock1 term -> true.
qa1_lock1 bot -> true.

qr2_lock2 acq_1 -> (1, qr2_lock2).
qr2_lock2 rel_1 -> (1, qr2_lock2).
qr2_lock2 acq_2 -> (1, qa2_lock2).
qr2_lock2 rel_2 -> false.
qr2_lock2 label_1 -> false.
qr2_lock2 label_2 -> false.
qr2_lock2 sp -> ((1, qr2_lock2) \/ (2, qr2_lock2) \/
    ((1, qr1_lock2) /\ (2, qr1_lock2))).
qr2_lock2 jo -> (1, qr2_lock2).
qr2_lock2 br -> true.
qr2_lock2 term -> false.
qr2_lock2 bot -> false.

qr1_lock2 acq_1 -> (1, qr1_lock2).
qr1_lock2 rel_1 -> (1, qr1_lock2).
qr1_lock2 acq_2 -> (1, qa1_lock2).
qr1_lock2 rel_2 -> false.
qr1_lock2 label_1 -> false.
qr1_lock2 label_2 -> false.
qr1_lock2 sp -> ((1, qr1_lock2) \/ (2, qr1_lock2)).
qr1_lock2 jo -> (1, qr1_lock2).
qr1_lock2 br -> true.
qr1_lock2 term -> false.
qr1_lock2 bot -> false.

qa2_lock2 acq_1 -> (1, qa2_lock2).
qa2_lock2 rel_1 -> (1, qa2_lock2).
qa2_lock2 acq_2 -> false.

qa2_lock2 rel_2 -> (1, qr2_lock2).
qa2_lock2 label_1 -> false.
qa2_lock2 label_2 -> false.
qa2_lock2 sp -> ((1, qa2_lock2) \/ (2, qr2_lock2) \/
    ((1, qa1_lock2) /\ (2, qr1_lock2))).
qa2_lock2 jo -> (1, qa2_lock2).
qa2_lock2 br -> true.
qa2_lock2 term -> false.
qa2_lock2 bot -> false.

qa1_lock2 acq_1 -> (1, qa1_lock2).
qa1_lock2 rel_1 -> (1, qa1_lock2).
qa1_lock2 acq_2 -> false.
qa1_lock2 rel_2 -> (1, qr1_lock2).
qa1_lock2 label_1 -> true.
qa1_lock2 label_2 -> true.
qa1_lock2 sp -> ((1, qa1_lock2) \/ (2, qr1_lock2)).
qa1_lock2 jo -> (1, qa1_lock2).
qa1_lock2 br -> false.
qa1_lock2 term -> true.
qa1_lock2 bot -> true.

qb_0 acq_1 -> (1, qb_0).
qb_0 rel_1 -> (1, qb_0).
qb_0 acq_2 -> (1, qb_0).
qb_0 rel_2 -> (1, qb_0).
qb_0 label_1 -> false.
qb_0 label_2 -> false.
qb_0 sp -> ((1, qb_0) \/ ((2, qb_0) \/ ((1, qb_jo) /\
    (2, qb_bot)))).
qb_0 jo -> (1, qb_0).
qb_0 br -> false.
qb_0 term -> false.
qb_0 bot -> false.

qb_jo acq_1 -> (1, qb_jo).
qb_jo rel_1 -> (1, qb_jo).
qb_jo acq_2 -> (1, qb_jo).
qb_jo rel_2 -> (1, qb_jo).
qb_jo label_1 -> false.
qb_jo label_2 -> false.
qb_jo sp -> (1, qb_jo).
qb_jo jo -> true.
qb_jo br -> false.
qb_jo term -> false.
qb_jo bot -> false.

qb_bot acq_1 -> (1, qb_bot).
qb_bot rel_1 -> (1, qb_bot).
qb_bot acq_2 -> (1, qb_bot).
qb_bot rel_2 -> (1, qb_bot).
qb_bot label_1 -> true.
qb_bot label_2 -> false.
qb_bot sp -> (1, qb_bot).
qb_bot jo -> (1, qb_bot).
qb_bot br -> false.
qb_bot term -> false.
qb_bot bot -> true.

qb_a_lock1 acq_1 -> (1, qb_a_lock1).
qb_a_lock1 rel_1 -> (1, qb_r_lock1).
qb_a_lock1 acq_2 -> (1, qb_a_lock1).
qb_a_lock1 rel_2 -> (1, qb_a_lock1).
qb_a_lock1 label_1 -> false.
qb_a_lock1 label_2 -> false.
qb_a_lock1 sp -> ((1, qb_a_lock1) \/ (2, qb_r_lock1) \/
    ((1, qb_keep_lock1) /\ (2, qb_use_lock1))).
qb_a_lock1 jo -> (1, qb_a_lock1).
qb_a_lock1 br -> false.
qb_a_lock1 term -> false.
qb_a_lock1 bot -> false.

qb_r_lock1 acq_1 -> (1, qb_a_lock1).
```

```
qb_r_lock1 rel_1 -> (1, qb_r_lock1).
qb_r_lock1 acq_2 -> (1, qb_r_lock1).
qb_r_lock1 rel_2 -> (1, qb_r_lock1).
qb_r_lock1 label_1 -> false.
qb_r_lock1 label_2 -> false.
qb_r_lock1 sp -> ((1, qb_r_lock1) \/ (2, qb_r_lock1)).
qb_r_lock1 jo -> (1, qb_r_lock1).
qb_r_lock1 br -> false.
qb_r_lock1 term -> false.
qb_r_lock1 bot -> false.

qb_use_lock1 acq_1 -> true.
qb_use_lock1 rel_1 -> (1, qb_use_lock1).
qb_use_lock1 acq_2 -> (1, qb_use_lock1).
qb_use_lock1 rel_2 -> (1, qb_use_lock1).
qb_use_lock1 label_1 -> false.
qb_use_lock1 label_2 -> false.
qb_use_lock1 sp -> ((1, qb_use_lock1) \/ ((1, qb_jo) /\
    (2, qb_use_lock1))).
qb_use_lock1 jo -> (1, qb_use_lock1).
qb_use_lock1 br -> false.
qb_use_lock1 term -> false.
qb_use_lock1 bot -> false.

qb_keep_lock1 acq_1 -> (1, qb_keep_lock1).
qb_keep_lock1 rel_1 -> false.
qb_keep_lock1 acq_2 -> (1, qb_keep_lock1).
qb_keep_lock1 rel_2 -> (1, qb_keep_lock1).
qb_keep_lock1 label_1 -> false.
qb_keep_lock1 label_2 -> false.
qb_keep_lock1 sp -> (1, qb_keep_lock1).
qb_keep_lock1 jo -> true.
qb_keep_lock1 br -> false.
qb_keep_lock1 term -> false.
qb_keep_lock1 bot -> false.

qb_a_lock2 acq_1 -> (1, qb_a_lock2).
qb_a_lock2 rel_1 -> (1, qb_a_lock2).
qb_a_lock2 acq_2 -> (1, qb_a_lock2).
qb_a_lock2 rel_2 -> (1, qb_r_lock2).
qb_a_lock2 label_1 -> false.
qb_a_lock2 label_2 -> false.
qb_a_lock2 sp -> ((1, qb_a_lock2) \/ (2, qb_r_lock2) \/
    ((1, qb_keep_lock2) /\ (2, qb_use_lock2))).
qb_a_lock2 jo -> (1, qb_a_lock2).
qb_a_lock2 br -> false.
qb_a_lock2 term -> false.
qb_a_lock2 bot -> false.

qb_r_lock2 acq_1 -> (1, qb_r_lock2).
qb_r_lock2 rel_1 -> (1, qb_r_lock2).
qb_r_lock2 acq_2 -> (1, qb_a_lock2).
qb_r_lock2 rel_2 -> (1, qb_r_lock2).
qb_r_lock2 label_1 -> false.
qb_r_lock2 label_2 -> false.
qb_r_lock2 sp -> ((1, qb_r_lock2) \/ (2, qb_r_lock2)).
qb_r_lock2 jo -> (1, qb_r_lock2).
qb_r_lock2 br -> false.
qb_r_lock2 term -> false.
qb_r_lock2 bot -> false.

qb_use_lock2 acq_1 -> (1, qb_use_lock2).
qb_use_lock2 rel_1 -> (1, qb_use_lock2).
qb_use_lock2 acq_2 -> true.
qb_use_lock2 rel_2 -> (1, qb_use_lock2).
qb_use_lock2 label_1 -> false.
qb_use_lock2 label_2 -> false.
qb_use_lock2 sp -> ((1, qb_use_lock2) \/ ((1, qb_jo) /\
    (2, qb_use_lock2))).
qb_use_lock2 jo -> (1, qb_use_lock2).
qb_use_lock2 br -> false.
qb_use_lock2 term -> false.
qb_use_lock2 bot -> false.

qb_keep_lock2 acq_1 -> (1, qb_keep_lock2).
qb_keep_lock2 rel_1 -> (1, qb_keep_lock2).
qb_keep_lock2 acq_2 -> (1, qb_keep_lock2).
qb_keep_lock2 rel_2 -> false.
qb_keep_lock2 label_1 -> false.
qb_keep_lock2 label_2 -> false.
qb_keep_lock2 sp -> (1, qb_keep_lock2).
qb_keep_lock2 jo -> true.
qb_keep_lock2 br -> false.
qb_keep_lock2 term -> false.
qb_keep_lock2 bot -> false.

q_fa_lock1 acq_1 -> (1, q_fa_lock1).
q_fa_lock1 rel_1 -> false.
q_fa_lock1 acq_2 -> (1, q_fa_lock1).
q_fa_lock1 rel_2 -> (1, q_fa_lock1).
q_fa_lock1 label_1 -> true.
q_fa_lock1 label_2 -> true.
q_fa_lock1 sp -> (1, q_fa_lock1).
q_fa_lock1 jo -> (1, q_fa_lock1).
q_fa_lock1 br -> false.
q_fa_lock1 term -> true.
q_fa_lock1 bot -> true.

q_acq_lock1 acq_1 -> true.
q_acq_lock1 rel_1 -> (1, q_acq_lock1).
q_acq_lock1 acq_2 -> (1, q_acq_lock1).
q_acq_lock1 rel_2 -> (1, q_acq_lock1).
q_acq_lock1 label_1 -> false.
q_acq_lock1 label_2 -> false.
q_acq_lock1 sp -> ((1, q_acq_lock1) \/ (2, q_acq_lock1))
    .
q_acq_lock1 jo -> (1, q_acq_lock1).
q_acq_lock1 br -> false.
q_acq_lock1 term -> false.
q_acq_lock1 bot -> false.

q_fa_lock2 acq_1 -> (1, q_fa_lock2).
q_fa_lock2 rel_1 -> (1, q_fa_lock2).
q_fa_lock2 acq_2 -> (1, q_fa_lock2).
q_fa_lock2 rel_2 -> false.
q_fa_lock2 label_1 -> true.
q_fa_lock2 label_2 -> true.
q_fa_lock2 sp -> (1, q_fa_lock2).
q_fa_lock2 jo -> (1, q_fa_lock2).
q_fa_lock2 br -> false.
q_fa_lock2 term -> true.
q_fa_lock2 bot -> true.

q_acq_lock2 acq_1 -> (1, q_acq_lock2).
q_acq_lock2 rel_1 -> (1, q_acq_lock2).
q_acq_lock2 acq_2 -> true.
q_acq_lock2 rel_2 -> (1, q_acq_lock2).
q_acq_lock2 label_1 -> false.
q_acq_lock2 label_2 -> false.
q_acq_lock2 sp -> ((1, q_acq_lock2) \/ (2, q_acq_lock2))
    .
q_acq_lock2 jo -> (1, q_acq_lock2).
q_acq_lock2 br -> false.
q_acq_lock2 term -> false.
q_acq_lock2 bot -> false.

q_tca_lock1_lock1 acq_1 -> ((1, q_tca_lock1_lock1) \/
    ((1, q_fa_lock1) /\ ((1, q_acq_lock1) \/ ((1,
    q_acq_lock2) /\ (1, q_tca_lock2_lock1))))).
q_tca_lock1_lock1 rel_1 -> (1, q_tca_lock1_lock1).
q_tca_lock1_lock1 acq_2 -> (1, q_tca_lock1_lock1).
q_tca_lock1_lock1 rel_2 -> (1, q_tca_lock1_lock1).
q_tca_lock1_lock1 label_1 -> false.
q_tca_lock1_lock1 label_2 -> false.
q_tca_lock1_lock1 sp -> ((1, q_tca_lock1_lock1) \/ (2,
    q_tca_lock1_lock1)).
```

```
q_tca_lock1_lock1 jo -> (1, q_tca_lock1_lock1).
q_tca_lock1_lock1 br -> false.
q_tca_lock1_lock1 term -> false.
q_tca_lock1_lock1 bot -> false.

q_tca_lock1_lock2 acq_1 -> ((1, q_tca_lock1_lock2) \/
    ((1, q_fa_lock1) /\ (1, q_acq_lock2))).
q_tca_lock1_lock2 rel_1 -> (1, q_tca_lock1_lock2).
q_tca_lock1_lock2 acq_2 -> (1, q_tca_lock1_lock2).
q_tca_lock1_lock2 rel_2 -> (1, q_tca_lock1_lock2).
q_tca_lock1_lock2 label_1 -> false.
q_tca_lock1_lock2 label_2 -> false.
q_tca_lock1_lock2 sp -> ((1, q_tca_lock1_lock2) \/ (2,
    q_tca_lock1_lock2)).
q_tca_lock1_lock2 jo -> (1, q_tca_lock1_lock2).
q_tca_lock1_lock2 br -> false.
q_tca_lock1_lock2 term -> false.
q_tca_lock1_lock2 bot -> false.

q_tca_lock2_lock1 acq_1 -> (1, q_tca_lock2_lock1).
q_tca_lock2_lock1 rel_1 -> (1, q_tca_lock2_lock1).
q_tca_lock2_lock1 acq_2 -> ((1, q_tca_lock2_lock1) \/
    ((1, q_fa_lock2) /\ (1, q_acq_lock1))).
q_tca_lock2_lock1 rel_2 -> (1, q_tca_lock2_lock1).
q_tca_lock2_lock1 label_1 -> false.
q_tca_lock2_lock1 label_2 -> false.
q_tca_lock2_lock1 sp -> ((1, q_tca_lock2_lock1) \/ (2,
    q_tca_lock2_lock1)).
q_tca_lock2_lock1 jo -> (1, q_tca_lock2_lock1).
q_tca_lock2_lock1 br -> false.
q_tca_lock2_lock1 term -> false.
q_tca_lock2_lock1 bot -> false.

q_tca_lock2_lock2 acq_1 -> (1, q_tca_lock2_lock2).
q_tca_lock2_lock2 rel_1 -> (1, q_tca_lock2_lock2).
q_tca_lock2_lock2 acq_2 -> ((1, q_tca_lock2_lock2) \/
    ((1, q_fa_lock2) /\ ((1, q_acq_lock2) \/ ((1,
    q_acq_lock1) /\ (1, q_tca_lock1_lock2))))).
q_tca_lock2_lock2 rel_2 -> (1, q_tca_lock2_lock2).
q_tca_lock2_lock2 label_1 -> false.
q_tca_lock2_lock2 label_2 -> false.
q_tca_lock2_lock2 sp -> ((1, q_tca_lock2_lock2) \/ (2,
    q_tca_lock2_lock2)).
q_tca_lock2_lock2 jo -> (1, q_tca_lock2_lock2).
q_tca_lock2_lock2 br -> false.
q_tca_lock2_lock2 term -> false.
q_tca_lock2_lock2 bot -> false.
```

## C. AUTOMATA PROOFS

Gawlitza et al. prove that an automaton that determines the four properties (i) double final acquisition, (ii) child termination, (iii) join-lock deadlocks, and (iv) lock-acquisition-cycle deadlocks determines schedulability [2]. We give the following constructions and prove that they meet each criterion respectively. Together with the cycle-detection automata in Section 5.1.1 these automata meet all the criteria for determining join-lock-sensitive schedulability. Thus when disjuncted together with the pairwise reachability automaton at the end of this section they form an automaton that rejects join-lock-sensitive schedulable trees that are pairwise reachable. This disjunction is correct because all the properties bar lock-acquisition cycle deadlock detection are true of all action sequences in a tree or none and thus if any action sequence is unschedulable then all of them are. That is to say although finding one action sequence with one of these properties naïvely doesn't imply that all action sequences are unschedulable, in all cases bar lock-acquisiton-cycle detection this does hold. In the case of lock-acquisiton-cycle detection, the automaton rejects only if all action sequences are unschedulable. Naïvely one might assume that if half the action sequences have lock-acquisiton-cycles, and the other half are unschedulable for some other reason the tree would not be detected as safe. However consider an action tree, $A$ with the action sequences $S$ that are partitioned into $S_A$ and $S_B$. Let all action sequences in $S_A$ have lock-acquisiton-cycles, and all action sequences in $S_B$ have double final acquisitions (DFAs), and further, let no action sequences in $S_B$ have a lock-acquisiton-cycle. Because the lock-acquisiton-cycle detecting automaton only considers a tree safe if all action sequences have lock-acquisiton-cycles, the lock-acquisition-cycle automaton would not consider $A$ safe, as some action sequences do not have lock-acquisition cycles. Thus it might be naïvely assumed that the tree could be improperly detected as schedulable, as potentially not all action sequences in $S_A$ have DFAs. However, because DFA is a tree property, all action sequences must have this property or not, thus, if $S_B$ is non-empty, then all action sequences, including those in $S_A$ have DFAs, and thus the tree is correctly detected as safe.

### C.1 Double Final Acquisition

A tree is said to have a DFA if there are two threads, $t_1, t_2$ such that $t_1 \neq t_2$, that terminate whilst holding the same lock. That is to say $\exists l \in Locks. \exists t_1, t_2 \in Threads. t_1 \neq t_2 \land t_1$ finally acquires $l \land t_2$ finally acquires $l$. Because there is a finite and fixed number of locks we can replace the $\exists$ with a disjunction over all the locks. For each lock, $x$, we create an instance of the DFA automaton $\mathcal{D}_x$

$\mathcal{D}_x$ is formed by the power construction of two smaller automata. A lock deterring automaton $\mathcal{FA}_x$ and a spawn finding automaton $\mathcal{S}$. Informally $\mathcal{FA}_x$ returns true if there is a thread that finally acquires $lock_x$. Formally $\mathcal{FA}_x :=  \langle \Sigma, \{qr_x, qa_x\}, qr_x, \delta_{\mathcal{FA}_x} \rangle$ where:

$$\Sigma = \{\langle sp \rangle : 2, \langle jo \rangle : 1, \langle \$ \rangle : 0, \langle \bot \rangle : 0\}$$
$$\cup \{\langle label \rangle : 0 \mid label \in Labels\}$$
$$\cup \{\langle acq_y \rangle : 1, \langle rel_y \rangle : 1 \mid y \in Locks\}$$
$$\text{and } \delta_{\mathcal{FA}_x} = \{qr_x \langle sp \rangle \to (1, qr_x) \land (2, qr_x),$$
$$qr_x \langle jo \rangle \to (1, qr_x),$$
$$qr_x \langle \$ \rangle \to false,$$
$$qr_x \langle \bot \rangle \to false\}$$
$$\cup$$
$$\{qr_x \langle label \rangle \to false \mid label \in Label\}$$
$$\cup$$
$$\{qr_x \langle acq_y \rangle \to (1, qr_x),$$
$$qr_x \langle rel_y \rangle \to (1, qr_x) \mid y \in Locks/x\}$$
$$\cup$$
$$\{qr_x \langle acq_x \rangle \to (1, qa_x),$$
$$qr_x \langle rel_x \rangle \to false\}$$
$$\cup$$
$$\{qa_x \langle sp \rangle \to (1, qa_x) \land (2, qr_x),$$
$$qa_x \langle jo \rangle \to (1, qa_x),$$
$$qa_x \langle \$ \rangle \to true,$$
$$qa_x \langle \bot \rangle \to true\}$$
$$\cup$$

$$\{qa_x \langle label \rangle \to true \mid label \in Label\}$$
$$\cup$$
$$\{qa_x \langle acq_y \rangle \to (1, qa_x),$$
$$qa_x \langle rel_y \rangle \to (1, qa_x) \mid y \in Locks/x\}$$
$$\cup$$
$$\{qa_x \langle acq_x \rangle \to false,$$
$$qa_x \langle rel_x \rangle \to (1, qr_x)\}$$

$\mathcal{S}$ returns true if there is a spawn such that on both the parent and child branch there is a terminating thread. Formally $\mathcal{S} := \langle \Sigma, \{q_2, q_1\}, q_1, \delta_\mathcal{S} \rangle$ where:

$$\delta_\mathcal{S} = \{q_2 \langle sp \rangle \to ((1, q_2) \vee (2, q_2)) \vee ((1, q_1) \wedge (2, q_1)),$$
$$q_2 \langle jo \rangle \to (1, q_2),$$
$$q_2 \langle \$ \rangle \to false,$$
$$q_2 \langle \bot \rangle \to false\}$$
$$\cup$$
$$\{q_2 \langle label \rangle \to false \mid label \in Label\}$$
$$\cup$$
$$\{q_2 \langle acq_y \rangle \to (1, q_2),$$
$$q_2 \langle rel_y \rangle \to (1, q_2) \mid y \in Locks\}$$
$$\cup$$
$$\{q_1 \langle sp \rangle \to (1, q_1) \wedge (2, q_1),$$
$$q_1 \langle jo \rangle \to (1, q_1),$$
$$q_1 \langle \$ \rangle \to true,$$
$$q_1 \langle \bot \rangle \to true\}$$
$$\cup$$
$$\{q_1 \langle label \rangle \to true \mid label \in Label\}$$
$$\cup$$
$$\{q_1 \langle acq_y \rangle \to (1, q_1),$$
$$q_1 \langle rel_y \rangle \to (1, q_1) \mid y \in Locks\}$$

To prove that $\mathcal{D}_x$ is correct we must show that $\mathcal{L}(\mathcal{FA}_x \cap \mathcal{S})$ accepts all trees with DFA and rejects those that do not. First we show that $\mathcal{FA}_x$ accepts only threads that finally acquire $lock_x$. We note that the only state that has any *true* outgoing edges is $qa_x$. Therefore only when terminating in state $qa_x$ can the tree be accepted[6]. Only three actions lead to a change of state, $qr_x \langle acq_x \rangle$, $qa_x \langle rel_x \rangle$, and $qa_x \langle sp \rangle$. The last of these, $qa_x \langle sp \rangle$, keeps the parent thread in the same state, and sends the child thread to $qr_x$. Because $qr_x$ is the start state, this means that all threads start in state $qr_x$. This means that any threads that terminate without performing an $\langle acq_x \rangle$ are rejected. Because all threads start with no locks, any thread that terminates without performing an $\langle acq_x \rangle$ does not finally acquire $lock_x$, because they never acquire the lock. Because we require trees to be lock-well-formed the pattern of lock acquisitions and releases within a thread must be a prefix of $(\langle acq_x \rangle \langle rel_x \rangle)*$, i.e. a lock must be acquired once, and then released once before beginning again. Thus if $\mathcal{FA}_x$ is in state $qa_x$ then $lock_x$ has been acquired exactly one more times than it has been released, and thus the thread holds the lock. Therefore if the thread terminates whilst $\mathcal{FA}_x$ is in state $qa_x$, the thread terminates holding the lock, and thus finally acquires the lock. If $\mathcal{FA}_x$ terminates in state $qa_x$ it returns true. If $\mathcal{FA}_x$ is in state $qr_x$ the thread has acquired and released the thread an equal number of times, and thus does not hold the lock. If $\mathcal{FA}_x$ terminates in state $qr_x$ then the lock has not been finally acquired because it is not held. If $\mathcal{FA}_x$ terminates in state $qr_x$ it returns false. Therefore $\mathcal{FA}_x$ returns true iff there is a thread that finally acquires $lock_x$.

---

[6]Because we are evaluting using trivial Büchi acceptance condition infinite paths are returned as true. However we can ignore infinite trees because the action tree forest will contain a finite prefix that will be correctly rejected.

Next we show that $\mathcal{S}$ only accepts trees with two terminating threads.

**Lemma 2.** *For any two distinct threads then there is exactly one $\langle sp \rangle$ such that the parent branch has one thread and the child branch has the other.*

*Proof.* Given that the input is a tree, for there to be two distinct threads there must be a $\langle sp \rangle$ that separates them. There can only be one $\langle sp \rangle$ that separates them because given a separating $\langle sp \rangle$ any $\langle sp \rangle$ that occurs higher up the tree must have them both on the same branch, as branches of a tree do not rejoin, by definition. Any $\langle sp \rangle$ that occurs after the separating $\langle sp \rangle$ will only have one of the threads, again as trees do not rejoin. Therefore the two threads have a single separating $\langle sp \rangle$. □

$\mathcal{S}$ is looking for two distinct terminating threads and by Lemma 2, those two threads are separated by a single $\langle sp \rangle$. On reaching a spawn in the tree $\mathcal{S}$ non-deterministically decides if the spawn is the separating spawn. If not it searches the parent and child branches for the separating spawn. If it is the separating spawn $\mathcal{S}$ checks both branches satisfy state $q_1$. $q_1$ accepts any tree with a terminating thread. Thus if both branches satisfy $q_1$ then there are two distinct terminating threads in the tree.

The intersection of $\mathcal{L}(\mathcal{FA}_x)$ with $\mathcal{L}(\mathcal{S})$ will accept any tree with two terminating branches that also finally acquire $lock_x$, which is the definition of DFA.

### C.2 Child Termination

A tree is considered unschedulable if it has a parent thread with a $\langle jo \rangle$ with a corresponding child thread that doesn't perform the terminate action $\langle \$ \rangle$. We detect this property using an automaton $\mathcal{T} := \langle \Sigma, \{qb_0, qb_{jo}, qb_{bot}\}, qb_0, \delta_\mathcal{T} \rangle$ where:

$$\Sigma = \{\langle sp \rangle : 2, \langle jo \rangle : 1, \langle \$ \rangle : 0, \langle \bot \rangle : 0\}$$
$$\cup \{\langle label \rangle : 0 \mid label \in Labels\}$$
$$\cup \{\langle acq_y \rangle : 1, \langle rel_y \rangle : 1 \mid y \in Locks\}$$

and

$$\delta_\mathcal{T} = \{qb_0 \langle sp \rangle \to ((1, qb_0) \vee (2, qb_0)) \vee ((1, qb_{jo}) \wedge (2, qb_{bot})),$$
$$qb_0 \langle jo \rangle \to (1, qb_0),$$
$$qb_0 \langle \$ \rangle \to false,$$
$$qb_0 \langle \bot \rangle \to false\}$$
$$\cup$$
$$\{qb_0 \langle label \rangle \to false \mid label \in Label\}$$
$$\cup$$
$$\{qb_0 \langle acq_y \rangle \to (1, qb_0),$$
$$qb_0 \langle rel_y \rangle \to (1, qb_0) \mid y \in Locks\}$$
$$\cup$$

$$\{qb_{jo} \langle sp \rangle \to (1, qb_{jo})$$
$$qb_{jo} \langle jo \rangle \to true,$$
$$qb_{jo} \langle \$ \rangle \to false,$$
$$qb_{jo} \langle \bot \rangle \to false\}$$
$$\cup$$
$$\{qb_{jo} \langle label \rangle \to false \mid label \in Label\}$$
$$\cup$$
$$\{qb_{jo} \langle acq_y \rangle \to (1, qb_{jo}),$$
$$qb_{jo} \langle rel_y \rangle \to (1, qb_{jo}) \mid y \in Locks\}$$
$$\cup$$
$$\{qb_{bot} \langle sp \rangle \to (1, qb_{bot}) \wedge (2, qb_{bot})$$
$$qb_{bot} \langle jo \rangle \to false,$$
$$qb_{bot} \langle \$ \rangle \to true,$$
$$qb_{bot} \langle \bot \rangle \to false\}$$
$$\cup$$
$$\{qb_{bot} \langle label \rangle \to false \mid label \in Label\}$$
$$\cup$$
$$\{qb_{bot} \langle acq_y \rangle \to (1, qb_{bot}),$$
$$qb_{bot} \langle rel_y \rangle \to (1, qb_{bot}) \mid y \in Locks\}$$

Using Lemma 2 we can see that there must be a $\langle sp \rangle$ that separates the parent thread with the $\langle jo \rangle$ from the non-terminating child. As it traverses the tree $qb_0$ non-deterministically decides if a spawn is the separating spawn. If it is not it searches the child and parent branches for the separating spawn. If it is then it checks the parent thread does indeed contain a $\langle jo \rangle$. $qr_{jo}$ ensures that scoping is obeyed by only searching the parent thread for the $\langle jo \rangle$. $qr_{jo}$ returns true if the parent thread has a $\langle jo \rangle$. $qr_{jo}$ returns false in all other circumstances[7]. $qb_{bot}$ similarly searches for threads that terminate without performing the terminate action, $\langle \$ \rangle$. In this case threads that are infinite also return true, which is required to make this check safe. Thus together $\mathcal{T}$ accepts runs where there is a parent thread that has a $\langle jo \rangle$, and a child thread that does not terminate in a terminate action.

## C.3 Join-Lock Interaction

The join-lock dependency automaton determines whether there is a dependency between a join and a lock. That is to say, there is a thread that acquires a lock, spawns a child thread that depends on the lock, and performs a join without releasing the lock. This is unschedulable, and thus safe, because the parent requires the child to terminate before it will release the lock and the child will not terminate until the lock is released. This is the only possible pattern for join-lock dependency because joins only affect children that occur before the $\langle jo \rangle$, and thus the spawn must occur before the join to be dependant on it. Further the lock acquisition must occur before the child is spawned, or there is a schedule whereby the child acquires and releases the lock before the parent. Finally the lock release must not occur before the $\langle jo \rangle$ because if the lock is released before the $\langle jo \rangle$ it can be acquired by the child thread before the child has to terminate. Thus the $\langle acq_x \rangle \prec \langle sp \rangle \prec \langle jo \rangle$.

---

[7] Again with the caveat on infinite threads as discussed in the section on DFA.

Using Lemma 2 we note there is a single spawn that separates the parent thread with the join from the child thread that depends on the lock. We split the tree into three sections, the section before the separating spawn, and the two sections after. We then need to evaluate three different properties, (i) that the section before the separating spawn holds $lock_x$, (ii) that the parent thread after the spawn has a join before it releases $lock_x$, and (iii) that the child thread depends on $lock_x$..

To evaluate these properties we construct three sub-automata, $\mathcal{JL}_x^1, \mathcal{JL}_x^2, \mathcal{JL}_x^3$. Together these form the automaton $\mathcal{JL}_x$, which accepts trees with the property above. $\mathcal{JL}_x^1$ is a slightly modified version of the lock acquisition automata used in the DFA construction to decide whether $lock_x$ has been acquired. We reject threads that terminate, as we are looking for paths that extend threads with the $\langle acq_x \rangle \ldots \langle sp \rangle$ pattern, and expand the $q_b a_x \langle sp \rangle$ rule to accept threads that terminate at the separating $\langle sp \rangle$[8]. Specifically $\mathcal{JL}_x^1 \coloneqq \langle \Sigma, \{q_b a_x, q_b r_x\}, \delta_{\mathcal{JL}_x^1}, q_b r_x \rangle$ where

$$\Sigma = \{\langle sp \rangle : 2, \langle jo \rangle : 1, \langle \$ \rangle : 0, \langle \bot \rangle : 0\}$$
$$\cup \{\langle label \rangle : 0 \mid label \in Labels\}$$
$$\cup \{\langle acq_y \rangle : 1, \langle rel_y \rangle : 1 \mid y \in Locks\}$$

and $\delta_{\mathcal{JL}_x^1} = \{q_b r_x \langle sp \rangle \to (1, q_b r_x) \wedge (2, q_b r_x),$
$$q_b r_x \langle jo \rangle \to (1, q_b r_x),$$
$$q_b r_x \langle \$ \rangle \to false,$$
$$q_b r_x \langle \bot \rangle \to false\}$$
$$\cup$$
$$\{q_b r_x \langle label \rangle \to false \mid label \in Label\}$$
$$\cup$$
$$\{q_b r_x \langle acq_y \rangle \to (1, q_b r_x),$$
$$q_b r_x \langle rel_y \rangle \to (1, q_b r_x) \mid y \in Locks/x\}$$
$$\cup$$
$$\{q_b r_x \langle acq_x \rangle \to (1, q_b a_x),$$
$$q_b r_x \langle rel_x \rangle \to false\}$$
$$\cup$$
$$\{q_b a_x \langle sp \rangle \to ((1, q_b a_x) \vee (2, q_b r_x)) \vee (\top),$$
$$q_b a_x \langle jo \rangle \to (1, q_b a_x),$$
$$q_b a_x \langle \$ \rangle \to false,$$
$$q_b a_x \langle \bot \rangle \to false\}$$
$$\cup$$
$$\{q_b a_x \langle label \rangle \to false \mid label \in Label\}$$
$$\cup$$
$$\{q_b a_x \langle acq_y \rangle \to (1, q_b a_x),$$
$$q_b a_x \langle rel_y \rangle \to (1, q_b a_x) \mid y \in Locks/x\}$$
$$\cup$$
$$\{q_b a_x \langle acq_x \rangle \to false,$$
$$q_b a_x \langle rel_x \rangle \to (1, q_b r_x)\}$$

The only trees that can be accepted are those that have a $\langle sp \rangle$ whilst $\mathcal{JL}_x^1$ is in state $q_b a_x$. By the same logic used in Appendix C.1, $\mathcal{JL}_x^1$ can only be in state $q_b a_x$ if the tree holds $lock_x$. Therefore $\mathcal{JL}_x^1$ accepts trees that reach a spawn whilst holding $lock_x$.

---

[8] In practice we disjunct the rule with $(1, q_b keep_x) \wedge (2, q_b use_x)$, but for the purposes of the proof we disjunct the rule with $\top$ and examine postfixes of the language accepted by $\mathcal{JL}_x^1$.

$\mathcal{JL}_x^2$ accepts trees that have a $\langle jo \rangle$ before they have an $\langle rel_x \rangle$. Formally $\mathcal{JL}_x^2 \coloneqq \langle \Sigma, \{q_b keep_x\}, \delta_{\mathcal{JL}_x^2}, q_b keep_x \rangle$ where

$$\begin{aligned}
\delta_{\mathcal{JL}_x^2} = \{&q_b keep_x \, \langle sp \rangle \to (1, q_b keep_x), \\
&q_b keep_x \, \langle jo \rangle \to true, \\
&q_b keep_x \, \langle \$ \rangle \to false, \\
&q_b keep_x \, \langle \bot \rangle \to false\} \\
&\cup \\
\{&q_b keep_x \, \langle label \rangle \to false \mid label \in Label\} \\
&\cup \\
\{&q_b keep_x \, \langle acq_y \rangle \to (1, q_b keep_x), \\
&q_b keep_x \, \langle rel_y \rangle \to (1, q_b keep_x) \mid y \in Locks/x\} \\
&\cup \\
\{&q_b keep_x \, \langle acq_x \rangle \to (1, q_b keep_x), \\
&q_b keep_x \, \langle rel_x \rangle \to false\}
\end{aligned}$$

$\mathcal{JL}_x^2$ only examines the parent thread, because both joins and releases that occur on child threads do not affect the parent thread, i.e. they are in a different scope. $\mathcal{JL}_x^2$ accepts on $\langle jo \rangle$, and rejects on $\langle rel_x \rangle$. Because only actions on the parent thread are relevant, the actions must be scheduled linearly, thus the first to occur on the parent thread is the first to be scheduled. Therefore any thread with a $\langle jo \rangle$ before a $\langle rel_x \rangle$ must schedule the $\langle jo \rangle$ before any (in scope) $\langle rel_x \rangle$ and vice versa[9]. This is the property required for $\mathcal{JL}_x$

$\mathcal{JL}_x^3$ accepts if the child thread depends on $lock_x$. A thread, $t$, is said to depend on $lock_x$ if it acquires it, or if one of its children acquires it and the child terminates[10]. If we drop the property that the child must terminate the child may never schedule the acquisition, and thus the property may not hold. Therefore to decide if $t$ depends on $lock_x$ each generation must require its child to terminate. Formally $\mathcal{JL}_x^3 \coloneqq \langle \Sigma, \{q_b use_x, q_b jo\}, \delta_{\mathcal{JL}_x^3}, q_b use_x \rangle$ where:

$$\begin{aligned}
\delta_{\mathcal{JL}_x^3} = \{&q_b use_x \, \langle sp \rangle \to (1, q_b use_x) \vee ((1, q_b jo) \wedge (2, q_b use_x)), \\
&q_b use_x \, \langle jo \rangle \to (1, q_b use_x), \\
&q_b use_x \, \langle \$ \rangle \to false, \\
&q_b use_x \, \langle \bot \rangle \to false\} \\
&\cup \\
\{&q_b use_x \, \langle label \rangle \to false \mid label \in Label\} \\
&\cup \\
\{&q_b use_x \, \langle acq_y \rangle \to (1, q_b use_x), \\
&q_b use_x \, \langle rel_y \rangle \to (1, q_b use_x) \mid y \in Locks/x\} \\
&\cup
\end{aligned}$$

---
[9]If the thread has neither a $\langle jo \rangle$ or a $\langle rel_x \rangle$, but continues infinitely it would under a trivial Büchi condition be accepted, however because of the construction of the action tree forest this case is correctly distinguished.
[10]Note this definition differs from the definition of one lock depending on another.

$$\begin{aligned}
\{&q_b use_x \, \langle acq_x \rangle \to true, \\
&q_b use_x \, \langle rel_x \rangle \to false\} \\
&\cup \\
\{&q_b jo \, \langle sp \rangle \to (1, q_b jo), \\
&q_b jo \, \langle jo \rangle \to true, \\
&q_b jo \, \langle \$ \rangle \to false, \\
&q_b jo \, \langle \bot \rangle \to false\} \\
&\cup \\
\{&q_b jo \, \langle label \rangle \to false \mid label \in Label\} \\
&\cup \\
\{&q_b jo \, \langle acq_y \rangle \to (1, q_b jo), \\
&q_b jo \, \langle rel_y \rangle \to (1, q_b jo) \mid y \in Locks\}
\end{aligned}$$

$\mathcal{JL}_x^3$ determines if $t$ depends on $lock_x$, if there is an $\langle acq_x \rangle$ on the parent thread then $\mathcal{JL}_x^3$ accepts, and the definition of dependence is met. The only method to ensure that all children terminate is using joins. If some child uses $lock_x$ then only if all of its ancestor threads ensure their children terminate can we guarentee the $\langle acq_x \rangle$ is scheduled[11]. Together $\mathcal{JL}_x^1, \mathcal{JL}_x^2, \mathcal{JL}_x^3$ determine the three properties defined at the start, and thus when constructed into the overall automaton $\mathcal{JL}_x$ determine join-lock dependence.

### C.4 Pairwise Reachability

The pairwise reachability automaton, $\mathcal{PR}_l$, detects whether label $l \in Label$ is pairwise reachable. Formally $\mathcal{PR}_l \coloneqq \langle \Sigma, \{q_2 l, q_1 l\}, q_2 l, \delta_{\mathcal{PR}_l} \rangle$ where:

$$\begin{aligned}
\delta_{\mathcal{PR}_l} = \{&q_2 l \, \langle sp \rangle \to ((1, q_2 l) \wedge (2, q_2 l)) \wedge ((1, q_1 l) \vee (2, q_1 l)), \\
&q_2 l \, \langle jo \rangle \to (1, q_2 l), \\
&q_2 l \, \langle \$ \rangle \to true, \\
&q_2 l \, \langle \bot \rangle \to true\} \\
&\cup \\
\{&q_2 l \, \langle label \rangle \to true \mid label \in Label\} \\
&\cup \\
\{&q_2 l \, \langle acq_y \rangle \to (1, q_2 l), \\
&q_2 l \, \langle rel_y \rangle \to (1, q_2 l) \mid y \in Locks\} \\
&\cup \\
\{&q_1 l \, \langle sp \rangle \to (1, q_1 l) \wedge (2, q_1 l), \\
&q_1 l \, \langle jo \rangle \to (1, q_1 l), \\
&q_1 l \, \langle \$ \rangle \to true, \\
&q_1 l \, \langle \bot \rangle \to true\} \\
&\cup \\
\{&q_1 l \, \langle label \rangle \to true \mid label \in Label\} \\
&\cup \\
\{&q_1 l \, \langle l \rangle \to false\} \\
&\cup \\
\{&q_1 l \, \langle acq_y \rangle \to (1, q_1 l), \\
&q_1 l \, \langle rel_y \rangle \to (1, q_1 l) \mid y \in Locks\}
\end{aligned}$$

---
[11]If all ancestor threads ensure termination but the $\langle acq_x \rangle$ in unschedulable for some other reason, this will be picked up by one of the other automata.

Informally $q_2 l$ determines whether all paths are label free. Using Lemma 2 we can see that for this to be false there must be a $\langle sp \rangle$ that separates two threads that do terminate in $\langle l \rangle$. Because we are looking for a negative property, i.e. there is not a label we invert the pattern used in the other automata. For the automaton to reject it asserts every thread pair is not pairwise reachable. Thus at each spawn it checks both that there are no later spawns that are pairwise reachable, but also that the spawn in question is not pairwise reachable. $q_1 l$ checks a thread for any occurance of the label, rejecting if it is found, thus the disjunction of $q_1 l$ over a parent and child returns false if both the parent and child contain the label. Thus the entire automata evaluates to false if there is a $\langle sp \rangle$ that separates two threads that can reach $l$.

## D. ACTION FOREST CONSTRUCTION

Take an action tree $T$ on which you wish to determine the pairwise reachability of two sets of nodes in the tree, $P^1, P^2$, i.e. that it is possible to reach a node from $P^1$ and a node from $P^2$ simultaneously. We define the function $Crop_{label}(T, P^1, P^2) = [T_{(1,1)} \ldots T_{(n,m)}]$ where $n$ is the number of nodes in $P^1$ and $m$ is the number of nodes in $P^2$. $T_{(i,j)} = T$ truncated at node $P_i^1$, which replaced with $l_1$, and also truncated at node $P_j^2$, which is replaced with $l_2$. We then define the branching function $br_k$ that takes $k$ trees and outputs a forest.

$br_k\ T_1 \ldots T_k = \langle br \rangle\ T_1\ (\langle br \rangle\ T_2\ (\ldots \langle br \rangle\ T_{k-1}\ T_k) \ldots)$

Let $T_b = br_{n \cdot m}(Crop_{label}(T, P^1, P^2))$. To account for each finite prefix of the thread for each instance of a single-child concurrency operator, $\langle acq \rangle, \langle rel \rangle, \langle jo \rangle$ we define a tree where the operator is replaced with $\bot$. Formally we define:

$Crop_{operator}\ T = [T_1 \ldots T_q]$ where $T_i = T$ truncated at the $i$th concurrency operator[12] replaced with $\bot$.

These trees are also joined into a single larger forest with $br_k$ The action tree forest is thus constructed as:

$br_q\ (Crop_{operator}\ (T_b))$

---

[12] For any linear ordering of concurrency operators in the tree.